\documentclass{amsart}
\usepackage{amssymb}
\usepackage{a4}
\begin{document}
\def\version{File W10v4i.tex Last Changed  Aug 12 by KK}
\def\beq{\begin{eqnarray}}
\def\eeq{\end{eqnarray}}
\newcommand{\nn}{\nonumber}
\def\rightheadline{\hfill\folio\hfill}
\def\leftheadline{\hfill\folio\hfill} 
\newtheorem{theorem}{Theorem}[section]
\newtheorem{lemma}[theorem]{Lemma}
\newtheorem{remark}[theorem]{Remark}
\newtheorem{definition}[theorem]{Definition}
\def\qedbox{\hbox{$\rlap{$\sqcap$}\sqcup$}}
\def\BBT{{\mathcal{B}_T({\mathcal{S}})}}
\def\BBR{{\mathcal{B}_R(S)}}
\def\tr{{\operatorname{Tr}}}
\makeatletter
  \renewcommand{\theequation}{%
   \thesection.\arabic{equation}}
  \@addtoreset{equation}{section}
 \makeatother
\title{Heat trace asymptotics defined by transfer boundary conditions}
\author{Peter Gilkey, Klaus Kirsten and Dmitri Vassilevich}
\begin{address}{
%PG: Mathematics Department, University of Oregon, Eugene OR 97403 USA\newline
%\phantom{...a}
PG, KK and DV: Max Planck Institute for Mathematics 
in the Sciences, 
Inselstrasse 22-26, 04103 Leipzig, Germany}
\end{address}
\begin{email}{gilkey@darkwing.uoregon.edu, klaus.kirsten@mis.mpg.de,\newline vassil@itp.uni-leipzig.de}
\end{email}
\begin{abstract} 
We compute the first 5 terms in the short-time heat trace asymptotics
expansion for an operator of Laplace type with transfer boundary conditions
using the functorial properties of these invariants.
\end{abstract}
\keywords{Laplace type operator, heat trace asymptotics,
transfer boundary conditions}
\subjclass{58J50}
\maketitle
\section{Introduction} Let $M:=(M^+,M^-)$ be a 
pair of compact smooth manifolds of dimension $m$ 
which have a common
smooth boundary $\Sigma:=\partial M^+=\partial M^-$. A structure $\Xi$ over $M$ will be
a pair of corresponding structures $\Xi:=(\Xi^+,\Xi^-)$ over the
manifolds
$M^\pm$. Let $g$ be a Riemannian metric on $M$; we assume henceforth that
$g^+|_\Sigma=g^-|_\Sigma$, but do not assume any matching condition on the normal derivatives.
Let $V$ be a smooth vector bundle over $M$; we do not assume any relationship between
$V^+|_\Sigma$ and $V^-|_\Sigma$; in particular, we can consider the situation when we have
$\dim V^+\ne\dim  V^-$. Let $D$ be an operator of Laplace type on $C^\infty(V)$. The operator $D$
determines a natural connection $\nabla$ and a natural $0^{th}$ order operator $E$ so that
\cite{bran90-15-245}:
$$
D =-\left (
%\operatorname{Tr}(\nabla)^2 +E\right).
g^{ij}\nabla_i \nabla_j + E\right) .
$$

Let the inward unit normals $\nu^\pm$ of
$\Sigma\subset M^\pm$ determine $\nu$; note that $\nu^+ =-\nu^-$. Assume given auxiliary
impedance matching terms
$\mathcal{S}=\{S^{++},S^{+-},S^{-+},S^{--}\}$ where
$S^{\varepsilon\varrho}:V^\varrho|_\Sigma\rightarrow V^\varepsilon|_\Sigma$.
The transfer
boundary operator
$\BBT$ is defined by:
\begin{equation}
\BBT\phi:=\left\{
     \left(\begin{array}{rr}
          \nabla_{\nu^+}^++S^{++}\qquad&S^{+-}\\
          S^{-+}\qquad&\nabla_{\nu^-}^-+S^{--}\end{array}\right)
      \left(\begin{array}{l}\phi^+\\\phi^-\end{array}\right)
\right\}\bigg|_\Sigma\,.\label{BBT}
\end{equation}
The terms $S^{+-}$ and $S^{-+}$ connect the structures on $M^+$ and $M^-$ and are crucial to our
investigation. These boundary conditions arise physically in heat 
transfer problems (see
to Carslaw and Jaeger
\cite{CaJa86}),  some problems of quantum mechanics
\cite{Albeverio}, and in conformal field theory \cite{BBDO02}. 
More on various spectral problems appearing in the string 
theory context can be
found in \cite{DVNaples}.

Let $D_\BBT$ be the associated realization of $D$
 with the boundary condition $\BBT\phi=0$.
Let $Q$ be a smooth endomorphism of $V$ which we use to localize the heat trace.
As $t\downarrow 0$, there is a complete asymptotic expansion with locally computable coefficients:
\begin{equation}
\tr_{L^2} \left(Qe^{-tD_{\BBT}} \right) \sim
\sum_{n\ge 0} a_n(Q,D,\BBT ) t^{(n-m)/2}.
\label{htr}
\end{equation}

In a formal limiting case $S^{++}-S^{-+}=S^{--}-S^{+-}\to\infty$
while $v=2(S^{++}+S^{+-})$ is kept finite
one arrives at transmittal boundary conditions: $\phi^+=\phi^-$,
$\nabla_{\nu^+}\phi^++\nabla_{\nu^-}\phi^-=v\phi^+$. The heat trace
asymptotics for these boundary conditions have been studied in
\cite{BV99,GKV01,Moss00}. Some other particular cases of the boundary
operator (\ref{BBT}) have been considered in \cite{BVFS01,Grosche}.

Let $R_{ijkl}$ be the components of the Riemann curvature tensor, let
$\Omega$ be the curvature of $\nabla$, 
and let the second fundamental forms $L^\pm$ of $\Sigma\subset M^\pm$ determine $L$. We let Roman
indices $i$, $j$, $k$, and $l$ range from $1$ to $m$ and index a local orthonormal frame for the
tangent bundle of $M$ and let Roman indices $a$,
$b$, $c$ range from $1$ to $m-1$ and index a local orthonormal frame for
the tangent bundle of $\Sigma$. We adopt the Einstein convention and sum
over repeated indices. Let $\tr^\pm$ be the fiber trace in $V^\pm$, let `;' denote multiple
covariant differentiation with respect to the Levi-Civita connection on $M$ and $\nabla$, and let
`:' denote multiple covariant differentiation with respect to the Levi-Civita connection of
$\Sigma$ and $\nabla$. Let $S=(S^{++},S^{--})$.

 Local formulae which decouple can be written in the following format:
\begin{definition}\label{defn1a}
Let $\mathcal{E}(\nabla^*R,\nabla^*E,\nabla^*\Omega)$ and
$\mathcal{F}(\nabla^*R,\nabla^*E,\nabla^*\Omega,\nabla^*L,\nabla^*S)$ be  local
invariants on $M$ and
$\partial M$, respectively. Set:
\begin{eqnarray*}
&&\textstyle\int_M\tr(\mathcal{E}):=
  \textstyle\int_{M^+}\tr^+(\mathcal{E}^+)+
  \textstyle\int_{M^-}\tr^-(\mathcal{E}^-),\\
&&\textstyle\int_{\partial M}\tr(\mathcal{F}):=
  \textstyle\int_{\partial M^+}\tr^+(\mathcal{F}^+)
   +\textstyle\int_{\partial M^-}\tr^-(\mathcal{F}^-)=
  \textstyle\int_{\Sigma}\{\tr^+(\mathcal{F}^+)+\tr^-(\mathcal{F}^-)\}.
\end{eqnarray*}
\end{definition}
\noindent What is crucial is that the invariants $\mathcal{E}^\pm$ and $\mathcal{F}^\pm$ involve
only structures on $M^\pm$. We illustrate these two
types in the following examples:
\begin{eqnarray*}
&&\textstyle\int_M\tr(QR_{ijji}E)
 =\textstyle\int_{M^+}\tr^+(Q^+R_{ijji}^+E^+)+
  \textstyle\int_{M^-}\tr^-(Q^-R_{ijji}^-E^-),\\
&&\textstyle\int_{\partial M}\tr(QSL_{aa})
  =\textstyle\int_{\partial M^+}\tr^+(Q^+S^{++}L_{aa}^+)
    +\int_{\partial M^-}\tr^-(Q^-S^{--}L_{aa}^-).\end{eqnarray*}
There are, however, invariants which intertwine the two structures 
and which do not decouple; for
example, the following invariant is a `mixed' invariant which 
measures the interactions of these
two structures:
$$\textstyle\int_\Sigma\{\tr^+(Q^+S^{+-}S^{-+})
+\tr^-(Q^-S^{-+}S^{+-})\}.$$

The main result of this letter is the following:

\begin{theorem}\label{thm1.2}
With transfer boundary conditions, we have that:
\begin{enumerate}
\item 
$a_0(Q,D,\BBT)=(4\pi)^{-m/2}\int_{M}\tr(Q)$.
\smallskip\item $a_1(Q,D,\BBT)=
     (4\pi)^{(1-m)/2}\frac14\int_{\partial M}\tr(Q)$.
\smallskip\item $a_2(Q,D,\BBT)=(4\pi)^{-m/2}\frac16
     \int_{M}\tr\{Q(R_{ijji}+6E)\}$
\smallbreak\quad
$+(4\pi)^{-m/2}\frac16\int_{\partial
M}\tr\{Q(2L_{aa}+12S)+3Q_{;\nu}\}$.
\smallskip\item $a_3(Q,D,\BBT)$
\smallbreak\quad$=
(4\pi)^{(1-m)/2}\frac1{384}
\int_{\partial M}
\tr\{Q(96E+16R_{ijji} -8R_{a\nu\nu a}$
$+13L_{aa}L_{bb}$\smallbreak\qquad\qquad$
+2L_{ab}L_{ab}+96SL_{aa}+192S^2$
$+Q_{;\nu}(6L_{aa}+96S)+24Q_{;\nu\nu})\}$
\smallbreak\quad
$+(4\pi)^{(1-m)/2}\frac1{384}\int_\Sigma\{
\tr^+(192Q^+S^{+-}S^{-+})+\tr^-(192Q^-S^{-+}S^{+-})\}.$
\smallskip\item
$a_4 (Q,D,\BBT )$\smallbreak\quad
$=(4\pi)^{-m/2}\frac1{360}\int_{M}\tr\{Q(60E_{;kk}+60 R_{ijji}E
+180E^{2}+30\Omega^2$
\smallbreak\qquad\qquad$+12 R_{ijji;kk} + 5R_{ijji}R_{kllk} -
2R_{ikjk}R_{iljl}+2 R_{ijkl}R_{ijkl})\}$
\smallbreak\quad
$+(4\pi)^{-m/2}\frac1{360}\int_{\partial M} \tr\{
      Q ( 240 E _{;\nu} + 42 R _{ijji;\nu} 
         +24 L_{aa:bb}+120 E L_{aa}$
\smallbreak\qquad\qquad
$  + 20 R_{ijji} L_{aa} 
 + 4 R _{a\nu a \nu } L_{bb} -12 R_{a\nu b\nu} L_{ab}
 + 4 R_{abcb} L_{ac} $
\smallbreak\qquad\qquad
$  + \frac{40}3 L_{aa} L_{bb} L_{cc} 
 +8
   L_{ab} L_{ab}  L_{cc} + \frac{32}3 L_{ab} 
L_{bc}  L_{ac} + 360(SE+ES)$
\smallbreak\qquad\qquad
 $ + 120 S R_{ijji}
 + 144 S
 L_{aa} L_{bb} 
  + 48 S L_{ab} L_{ab}+ 480 S^2 L_{aa} + 480 S^3 $
\smallbreak\qquad\qquad
$
   + 120 S_{:aa} )
+ Q_{;\nu} (180 E + 30 R_{ijji}
+ 12 L_{aa} L_{bb} + 12
      L_{ab}  L_{ab} $
\smallbreak\qquad\qquad
$+ 72 S L_{aa} + 240 S^2)+  Q_{;\nu \nu } ( 24 L_{aa} 
+ 120 S ) + 30 Q_{;ii\nu}\} $
\goodbreak\smallbreak\qquad
$+(4\pi)^{-m/2}\frac1{360}\int_{\Sigma}
  \tr^+\{480(Q^+S^{++}+S^{++}Q^+)S^{+-}S^{-+}$
\smallbreak\qquad\qquad
$
 +480Q^+S^{+-}S^{--}S^{-+} $
\smallbreak\qquad\qquad
$ + (288 Q^+ L_{aa}^+ + 192 Q^+L_{aa}^- +240 Q^+_{;\nu^+}) 
  S^{+-} S^{-+}\}$
\smallbreak\qquad
$+(4\pi)^{-m/2}\frac1{360}\int_{\Sigma}
  \tr^-\{480(Q^-S^{--}+S^{--}Q^-)S^{-+}S^{+-}
$
\smallbreak\qquad\qquad
$
 +480Q^-S^{-+}S^{++}S^{+-}$
\smallbreak\qquad\qquad
$ + (288 Q^- L_{aa}^- + 192 Q^-L_{aa}^+ +240 Q^-_{;\nu^-}) 
  S^{-+} S^{+-}\}$.
\end{enumerate}
\end{theorem}

We may decompose the heat trace invariants in the form:
\begin{equation}\label{eqn1.3}
a_n(Q,D,\BBT)=a_n^M(Q,D)+a_n^{\partial
M}(Q,D,S)+a_n^\Sigma(Q,D,\mathcal{S}).
\end{equation}
The invariants $a_n^M$ and $a_n^{\partial M}$ decouple and can be expressed as local integrals of
the form given in Definition \ref{defn1a}; the invariant
$a_n^\Sigma$ involves integrals of mixed structures. Theorem \ref{thm1.2} reflects this
decomposition. We shall prove Theorem \ref{thm1.2} by analyzing the 3 terms appearing in Equation
(\ref{eqn1.3}) separately. Here is a brief guide to the remainder of this letter. In Section
\ref{Sect2}, we apply results of Branson and Gilkey \cite{bran90-15-245} concerning
the heat trace asymptotics with Robin boundary conditions to determine
$a_n^M$ and
$a_n^{\partial M}$. In Section
\ref{Sect3}, we express $a_n^\Sigma$ in terms of certain invariants with universal
undetermined coefficients (see Lemma \ref{lem3.1}); these new terms which measure the interaction
between the structures on $M^\pm$ are the heart of the matter. 
The proof of Theorem
\ref{thm1.2} is then completed in Sections \ref{Sect4} and \ref{Sect5} 
by determining the
universal coefficients of Lemma \ref{lem3.1}. In Section
\ref{Sect4}, we derive a new functorial property by doubling the manifold; 
in Section \ref{Sect5}, we use conformal variations. We refer
to \cite{gilk01v} for an analogous computation of the heat content asymptotics with transfer
boundary conditions.

\section{Robin boundary conditions}\label{Sect2}
Let $D$ be an operator of Laplace type on a compact Riemannian
manifold $N$ with smooth boundary $\partial N$ and let $S$ be an auxiliary endomorphism defined
on the boundary. Robin boundary conditions are defined by the operator:
$$\mathcal{B}_R(S)\phi:=(\nabla_\nu\phi+S\phi)|_{\partial N}.$$

If we take
$S^{+-}=0$ and $S^{-+}=0$, then the boundary conditions decouple so
\begin{eqnarray*}
   a_n(Q,D,\BBT)&=&a_n(Q^+,D^+,\mathcal{B}_R(S^{++}))+
   a_n(Q^-,D^-,\mathcal{B}_R(S^{--}))\\
  &=&a_n(Q,D,\BBR).\end{eqnarray*}
Thus we may use Branson-Gilkey-Vassilevich \cite{BGV97} (Theorem 4.1) 
to determine the invariants
$a_n^M(Q,D)$ and $a_n^{\partial M}(Q,D,S)$.
Furthermore, we see that all
the terms in the mixed integrals defining $a_n^\Sigma(Q,D,\BBT)$ must
contain either $S^{+-}$ or $S^{-+}$ and hence, since we are taking traces and have not identified
$V^+$ with $V^-$, both of these terms must appear in every mixed monomial as these are the only
structures relating $M^+$ to $M^-$.

As the boundary integrands describing $a_n^\Sigma$ are homogeneous of weight
$n-1$ and as the variables $S^{**}$ have weight 1, monomials which
contain both $S^{+-}$ and $S^{-+}$ have weight at least $2$ and thus do not
appear in the expansion of $a_n$ for $n\le2$. This completes the proof of
Theorem
\ref{thm1.2} (1)-(3).

\section{The mixed invariants}\label{Sect3}
We can identify the
general form of the invariants $a_n^\Sigma$ for $n\le4$ as follows:
 \begin{lemma}\label{lem3.1} There exist universal constants so
that:\begin{enumerate}
\item $a_{3}^\Sigma(Q,D,\BBT)$\smallbreak
\quad$= (4\pi )^{-m/2}\frac1{384}
\int_\Sigma \alpha_0\{
\tr^+ (Q^+S^{+-}S^{-+})+\tr^-(Q^-S^{-+}S^{+-})\}.$
\medbreak\item
$a_4^\Sigma (Q,D,\BBT )= (4\pi )^{-m/2}\frac1{360} 
\int_\Sigma\{$\smallbreak\quad
$\frac12c_1\tr^+(Q^+S^{++}S^{+-}S^{-+})
+\frac12c_1\tr^-(Q^-S^{--}S^{-+}S^{+-})$
\smallbreak\quad
$+\frac12c_2\tr^+(S^{++}Q^+S^{+-}S^{-+})
+\frac12c_2\tr^-(S^{--}Q^-S^{-+}S^{+-})$
\smallbreak\quad
$+\alpha_2\tr^+(S^{++}S^{+-}Q^-S^{-+})+\alpha_2\tr^-( S^{--}S^{-+}Q^+S^{+-})$
\smallbreak\quad
$+\alpha_3L_{aa}^+\tr^+(Q^+S^{+-}S^{-+})+\alpha_3L_{aa}^-\tr^-(Q^-S^{-+}S^{+-})$
\smallbreak\quad
$+\alpha_4L_{aa}^-\tr^+(Q^+S^{+-}S^{-+})+\alpha_4L_{aa}^+\tr^-(Q^-S^{-+}S^{+-})$
\smallbreak\quad
$+\alpha_5\tr^+(Q^+_{;\nu^+}S^{+-}S^{-+})+
\alpha_5\tr^-(Q^-_{;\nu^-}S^{-+}S^{+-})
\}$ .
\smallskip\item $c_1=c_2$.
\end{enumerate}
\end{lemma}

\begin{proof} We observe first that the heat trace coefficient must be symmetric
with respect to interchanging the labels ``$+$'' and ``$-$''. Since we have written down a
complete basis of invariants of weight $2$ and $3$ which contain both $S^{-+}$ and $S^{+-}$,
assertions (1) and (2) now follow.

We generalize an argument from \cite{bran90-15-245} to prove assertion (3). If
$D$,
$Q$ and
$S^{**}$ are real, then $\tr \left( Qe^{-tD} \right)$ is real.
This shows that all universal constants given above are real.
Suppose now that the bundles $V^\pm$ are 
equipped with Hermitian inner products and that the operators
$D^\pm$ are formally self-adjoint. This means that the associated
connections $\nabla^\pm$ are unitary and the endomorphisms $E^\pm$
are symmetric. Suppose that $S^{++}$ and $S^{--}$ are self-adjoint,
and that $S^{+-}$ is the adjoint of $S^{-+}$. It then follows that $D$ is self-adjoint.
Therefore, 
$\tr \left( Qe^{-tD} \right)$ is real; this implies necessarily that $c_1=c_2$.
\end{proof}

We remark in passing that it is exactly this argument which shows that the term
$\int_M\tr(720QSE)$ appearing in \cite{bran90-15-245} for scalar $Q$ must be replaced by the term
$\int_M\tr(360Q(SE+ES))$ for endomorphism valued $Q$ \cite{BGV97}.

\medbreak Since $c_1=c_2$, the lack of commutativity involved in
dealing with endomorphisms
plays no role; thus it suffices to consider the scalar case where everything is commutative. We
assume therefore for the remainder of this letter that the bundles $V^\pm=M^\pm\times\mathbb{C}$
are trivial line bundles and that the operators
$D^\pm$ are scalar. Thus we may drop `$\tr$' from the notation. We set
$\alpha_1:=c_1=c_2$ - the symmetrization term then becomes
$$(4\pi)^{-m/2}\textstyle\frac1{360}\int_\Sigma
\alpha_1(Q^+S^{++}S^{+-}S^{-+}+Q^-S^{--}S^{-+}S^{+-}).$$

\section{Doubling the manifold}\label{Sect4}

In Section \ref{Sect2}, we related the heat trace asymptotics for transfer and Robin boundary
conditions by taking
$S^{+-}=S^{-+}=0$. We now give a different relationship between transfer and Robin boundary
conditions related to doubling the manifold.

\begin{lemma}\label{lem4.1} Let $M^\pm:=M^0$ be a $m$-dimensional 
Riemannian manifold with boundary $\partial M^0 = \Sigma$ 
and let $D^\pm = D^0$ be a scalar operator of Laplace type. Fix an angle
$0<\theta<\frac\pi2$. Let $S^{++}$ and $S^{+-}$ be arbitrary. Set:
$$\begin{array}{ll}
S^{-+}&:= S^{+-},\\
S^{--}&:= S^{++} + (\tan \theta -\cot \theta)\phantom{.} S^{+-},\\
S_\phi&:=S^{++}+\tan\theta \phantom{.}S^{+-}=S^{--}+\cot\phantom{.}\theta\phantom{.}
S^{-+},\\ 
S_\psi&:=S^{++}-\cot\phantom{.}\theta S^{+-}=S^{--}-\tan\phantom{.}\theta\phantom{.} S^{-+}.
\end{array}$$
Then:
\begin{eqnarray*}
a_n(Q,D,\BBT)&=&
a_n(\cos^2\theta\phantom{.} Q^++\sin^2\theta\phantom{.} Q^-,D^0,\mathcal{B}_{R(S_\phi)})\\
&+&a_n(\sin^2\theta\phantom{.} Q^++\cos^2\theta\phantom{.} Q^-,D^0,\mathcal{B}_{R(S_\psi)}).
\end{eqnarray*}
\end{lemma}

\begin{proof} If $u,v\in C^\infty(V^0)$, define $u^\phi,v^\psi\in C^\infty(M)$ by setting
$$\begin{array}{ll}
   u^\phi(x^+)=\phantom{-}\cos\theta\phantom{.}u(x),&u^\phi(x^-)=\sin\theta\phantom{.}u(x)\\
   v^\psi(x^+)=-\sin\theta\phantom{.}v(x),&v^\phi(x^-)=\cos\theta\phantom{.}v(x).
\end{array}$$
The conditions
$\BBT u^\phi=0$ and $\BBT v^\psi=0$ are equivalent to the conditions:
$$\begin{array}{ll}
(\nabla_{\nu^0}+S^{++}+\tan\theta\phantom{.} S^{+-})u|_{\partial M^0}=0,&
(\nabla_{\nu^0}+S^{--}+\cot\theta\phantom{.} S^{-+})u|_{\partial M^0}=0,\\ 
(\nabla_{\nu^0}+S^{++}-\cot\theta\phantom{.} S^{+-})v|_{\partial M^0}=0,&
(\nabla_{\nu^0}+S^{--}-\tan\theta\phantom{.} S^{-+})v|_{\partial M^0}=0,
\end{array}
$$
or equivalently to the conditions
$(\nabla_{\nu^0}+S_\phi)u|_{\partial M^0}=0$ and 
$(\nabla_{\nu^0}+S_\psi)v|_{\partial M^0}=0$.

\medbreak Let $\{\lambda_i,u_i\}$ and $\{\mu_j,v_j\}$ be discrete spectral resolutions
for
$D^0$ for Robin boundary conditions $\mathcal{B}_R(S_\phi)$ and 
$\mathcal{B}_R(S_\phi)$. Since
$$
\begin{array}{llll}
Du^\phi_i=\lambda_iu^\phi_i,&Dv^\psi_j=\mu_jv^\psi_j,&
\BBT u^\phi_i=0,&\text{and }\BBT v^\psi_j=0,\end{array}$$and
since $\{u^\phi_i,v^\psi_j\}$ is a complete orthonormal basis for
$L^2(M)$,
$\{\lambda_i,u^\phi_i\}\cup\{\mu_j,v^\psi_j\}$ is a
discrete spectral resolution of $D$ with transfer 
boundary conditions $\BBT$. Thus we may compute:
\goodbreak
\medbreak
\quad$\textstyle\tr_{L^2}(Qe^{-tD_\BBT})=
\textstyle\int_{M}\textstyle\sum_iQe^{-t\lambda_i}|u_i^\phi|^2
+\int_M\sum_jQe^{-t\mu_j}|v_j^\psi|^2\}$
\medbreak\qquad\qquad
$=\textstyle\int_{M^0}\sum_i(\cos^2\theta\phantom{.}Q^++\sin^2\theta\phantom{.}Q^-)
|u_i|^2e^{-t\lambda_i}$
\medbreak\qquad\qquad
$+\textstyle\int_{M^0}\sum_j(\sin^2\theta\phantom{.}Q^++\cos^2\theta\phantom{.}Q^-)
|v_j|^2e^{-t\mu_j}$
\medbreak\qquad\qquad
$=\textstyle\tr_{L^2}(\cos^2\theta\phantom{.}Q^++\sin^2\theta\phantom{.}Q^-)
   e^{-tD^0_{\mathcal{B}_R(\phi)}}$
\medbreak\qquad\qquad
$+\textstyle\tr_{L^2}(\sin^2\theta\phantom{.}Q^++\cos^2\theta\phantom{.}Q^-)
   e^{-tD^0_{\mathcal{B}_R(\psi)}}$.
\end{proof}

We use Lemma \ref{lem4.1} as follows. We set $Q^-=0$. 
(The case $Q^- \neq 0$ may be used as a check, but no additional information is obtained.) We use
\cite{bran90-15-245} (Theorem 1.2), Lemma
\ref{lem3.1}, and Lemma \ref{lem4.1} to derive the following relations,
\medbreak\qquad\quad $
192 Q^+ (\cos^2 \theta S_\phi^2 + \sin^2 \theta S_\psi^2 )= 
192 Q^+ ( S^{++}S^{++} +S^{+-}S^{+-} ) $\medbreak\quad\qquad\qquad
$= 192 Q^+ S^{++}S^{++} + \alpha_0 Q^+ S^{+-}S^{+-}$,
\medbreak\qquad\quad
$480 Q^+ (\cos^2 \theta S_\phi^3  + \sin^2 \theta S_{\psi }^3 )$
\smallbreak\qquad\qquad\quad$=
    480 Q^+ (S^{++}S^{++}S^{++} + 3 S^{++} S^{+-}S^{+-} + S^{+-}S^{+-}S^{+-} [
    \tan \theta - \cot \theta ] )$
\smallbreak\quad\qquad\qquad$=480 Q^+S^{++}S^{++}S^{++} + \alpha _1 Q^+ S^{++}S^{+-}S^{+-}$
\smallbreak\qquad\qquad\qquad$+ \alpha _2 Q^+ 
          [ S^{++} + S^{+-} ( \tan \theta -\cot \theta ) ]S^{+-}S^{+-} ,$
\medbreak\qquad\quad $
          480 Q^+ L_{aa}( \cos^2 \theta S_\phi^2 + \sin^2 \theta 
           S_\psi^2 ) =480 Q^+ L_{aa} (S^{++}S^{++} + S^{+-}S^{+-})$
\smallbreak\qquad\qquad\quad$
     = 480 Q^+ L_{aa} S^{++}S^{++} + (\alpha _3 + \alpha _4 ) 
     Q^+ L_{aa}S^{+-}S^{+-},$
\smallbreak\qquad\quad $240 Q^+_{;\nu^+}  (\cos^2 \theta S_\phi^2 + \sin^2 \theta S_\psi^2 ) 
    = 240 Q^{+}_{;\nu^+} (S^{++}S^{++} +S^{+-}S^{+-} )$
\smallbreak\quad\qquad\qquad
$= 240 Q^{+}_{;\nu^+}S^{++}S^{++} + \alpha _5 Q^+_{;\nu _+}S^{+-}S^{+-}$.
\medbreak\noindent
This implies that:
\begin{equation}\label{eqn4.1}
192 = \alpha_0,\quad960=\alpha_1,\quad480 = \alpha _2,\quad
480 = \alpha _3 + \alpha _4, \quad \alpha_5 = 240. \end{equation}

\section{Conformal variations}\label{Sect5}

The missing information about $\{\alpha_3, \alpha_4\}$ is obtained via 
conformal transformations. As before, we deal only with the scalar situation. Given $(M,D)$ and
$\psi^+\in C^\infty(M^+)$, we vary the structures on
$M^+$ to define the one-parameter family of  operators 
$$D(\varepsilon):=(e^{2\varepsilon \psi^+} D^+,D^-)$$
with associated structures $g^+(\varepsilon):=e^{2\varepsilon\psi^+}g^+$,
$\nabla^+(\varepsilon )$, and $E^+(\varepsilon)$. To ensure that
$g^+(\varepsilon)|_\Sigma=g^-|_\Sigma$, we assume $\psi^+$ vanishes on $\Sigma$. Let
$Q=(Q^+,0)$, $\psi:=(\psi^+,0)$, and
$$\mathcal{S}(\varepsilon):=\mathcal{B}_T(\mathcal{S}(0))-\nabla(\varepsilon)\text{Id}.$$

The following Lemma is a purely formal computation; see \cite{bran90-15-245} for
details.

\begin{lemma}\label{lem5.1} Adopt the notational conventions established above. Then
\begin{enumerate}
\item $\partial_\varepsilon|_{\varepsilon =0} a_n 
 (1, D(\varepsilon ) ,\mathcal{B}_T(\mathcal{S}(\varepsilon))) = (m-n) a_n (\psi, D, \BBT )$.
\smallskip\item $\partial_\varepsilon|_{\varepsilon =0} a_n 
(e^{-2\varepsilon \psi } Q , D(\varepsilon ),\mathcal{B}_T(\mathcal{S}(\varepsilon))) =0 
\text{ for }m=n+2$.
\end{enumerate}
\end{lemma}

We use the following relations to apply Lemma \ref{lem5.1}:
$$
\begin{array}{ll}
\partial|_{\varepsilon =0} S^{++} (\varepsilon ) =\frac{m-2} 2 \psi^+ _{;\nu^+},&
S^{+-} (\varepsilon )=S^{+-}(0),\\
S^{-+} (\varepsilon )=S^{-+}(0),& 
S^{--}(\varepsilon )=S^{--}(0),\\
\partial_\varepsilon|_{\varepsilon =0} L_{aa}^+ (\varepsilon ) = -(m-1) \psi^+_{;\nu^+},&
\partial_\varepsilon|_{\varepsilon =0}\{\nabla^+_{\nu^+}
 (\varepsilon )( e^{-2\varepsilon \psi } Q)\} = -2Q \psi ^+_{; \nu^+}.
\end{array}$$

Clearly Lemma \ref{lem5.1} (1) yields no new information
as the localizing function is continuous on $\Sigma$ and thus cannot
separate the contributions from $\alpha _3$ and $\alpha_4$. In fact, 
comparing the coefficient of the invariant $\psi^+_{;\nu^+}S^{+-}S^{-+}$, one obtains
\beq
\textstyle\frac {m-2} 2 (\alpha _1 + \alpha _2 ) - (m-1) (\alpha _3 + \alpha _4 ) 
  = (m-4) \alpha _5 \nn
\eeq
which is consistent with Equation (\ref{eqn4.1}). However, Lemma \ref{lem5.1} (2) with $m=6$
yields the additional relation:
\beq
2\alpha_1 -5\alpha_3 -2\alpha_5 =0.\nn
\eeq
We use Equation (\ref{eqn4.1}) to complete the proof of Theorem \ref{thm1.2}
by computing:
\beq
\alpha_3 = 288, \quad \alpha_4 = 192.
\eeq

\section*{Acknowledgements}
The research of P.G.\ was partially supported by the NSF (USA)
and the MPI (Leipzig), K.K.\ and D.V.\ were supported by the MPI
(Leipzig).

\end{document}